# Active Galactic Nuclei as seen by the Spitzer Space Telescope

Mark Lacy[1] and Anna Sajina[2]

**The Spitzer Space Telescope revolutionized studies of Active Galactic Nuclei (AGNs). Its combined sensitivity and mapping speed at mid-infrared wavelengths revealed a substantial population of highly-obscured AGNs. This population implies a higher radiative accretion efficiency, and thus possibly a higher spin for black holes than indicated by surveys in the optical and X-ray. The unique mid-infrared spectrographic capability of Spitzer gave important insights into the distribution and nature of the dust surrounding AGNs, enabling the separation of AGN and starburst components, the detection of silicate features in emission from hot dust, and the identification of shocked gas associated with AGN activity. The sensitivity of Spitzer allowed almost complete identification of the host galaxies of samples of AGNs selected in the X-ray and radio. As we look forward to the James Webb Space Telescope, the lessons learned from Spitzer studies will inform observational programs with new and upcoming infrared facilities.**

Our understanding of mid-infrared emission from AGNs began with the use of early infrared instrumentation on ground-based telescopes in the late 1960s,[1,2] and was greatly expanded in the era of the Infrared Astronomy Satellite (IRAS) mission in the 1980s. Based on these early studies it was realized that AGNs had an excess of emission in the mid-infrared compared to normal galaxies.[3,4,5] The first physical scenario presented to explain infrared emission from AGNs was in the form of an evolutionary scenario.[6] In this picture, an AGN is formed after a galaxy merger event and subsequent nuclear starburst. The AGN is initially highly obscured by dust, and reprocessing of UV radiation from the accretion disk by this dust results in strong mid-infrared emission. The dust is eventually blown out by winds associated with the AGN, resulting in an unobscured (Type-1) quasar. This theory has continued to remain popular to this day, with the winds also possibly playing a role in disrupting star formation in the host galaxies through a feedback mechanism[7,8] (or helping to trigger star formation in some circumstances[9]).

By the time of the Infrared Space Observatory (ISO) mission, in the mid-1990s, the concept of the "Unified Scheme" of AGNs had become established.[10] This was inspired by the observation of broad emission lines in the polarized optical light of some Type-2 Seyfert galaxies, which only show narrow emission lines in their total intensity spectra. A toroidal structure of dust and gas around the nucleus was hypothesized to explain this, with Type-1 objects being seen along a line of sight perpendicular to the plane of the torus, and Type-2 objects in the plane of the torus, which blocks the direct view of the AGN. Further support for this orientation-based understanding of AGNs came from studies of radio galaxies and radio quasars, where evidence of relativistic beaming of radio jets is seen only in the quasar population.[11,12] It was soon realized that this torus could be the origin of the strong 3-20 μm mid-infrared emission seen in AGNs[13-16] (Figure 1). The interpretation of the predominant component of the obscured population of AGNs in terms of either being a population viewed at a particular set of orientations to the observer, or being AGNs in an early phase of their lifecycle, remains a topic of debate.

ISO provided mid-infrared spectra for a wide range of Galactic environments as well as bright nearby galaxies, allowing us to begin understanding the dependence of the mid-infrared spectra on the conditions of the ISM and its heating mechanisms. The improved resolution of ISO compared to IRAS allowed better discrimination of infrared emission from AGNs versus star formation in the non-nuclear parts of the host galaxy, and confirmed the existence of nuclear emission from hot dust.[18-22] These observations showed that the strength of prominent mid-infrared emission features arising from Polycyclic Aromatic Hydrocarbons (PAHs; large molecules containing multiple aromatic rings of carbon atoms) compared with the continuum can be used as a proxy for the relative role of star-formation to AGNs in the mid-infrared.

---

[1] National Radio Astronomy Observatory, Charlottesville, VA, USA
[2] Tufts University, Medford, MA, USA

The advent of Spitzer, with its superior resolution, sensitivity and spectroscopic capability enabled observations to build upon these earlier efforts and address several fundamental questions about the nature of AGNs. Questions relating to how AGNs are triggered, how AGN activity is related to star formation, the nature of AGNs obscured by dust and gas, and how the AGN phenomenon depends on accretion rate onto the central black hole were all informed by Spitzer observations. The ease with which Spitzer data found heavily obscured AGNs that are only detectable in the deepest X-ray fields (see Sections 2 and 4) made it a powerful tool for understanding the early stages of the evolution of individual AGNs. The impact of Spitzer for AGN studies has been huge: the Astrophysics Data Service lists 824 refereed papers whose abstract includes "Spitzer" and "AGN"; these papers have been cited 43,134 times. In a short review such as this one we cannot hope to do full justice to the range and depth of AGN science performed with *Spitzer*, but instead have chosen to focus on a few areas in which we feel Spitzer made key contributions, focusing on results from the larger Spitzer surveys and spectroscopic studies of AGN dust emission. A general discussion of dust in galaxies appears in ref.[23] and a discussion of PAH features in AGNs are discussed elsewhere in this issue[24] so we have omitted detailed discussion of AGN dust and PAH physics from this Review.

The three different instruments aboard Spitzer each informed studies of AGNs in different ways. The Infrared Array Camera (IRAC) instrument[25] imaged in four bands, centered on 3.6, 4.5, 5.8 and 8.0 $\mu$m (the 3.6 and 4.5 $\mu$m bands remained working after the cryogen was exhausted in 2009). Offering imaging at ~2 arcsecond resolution, and a 5-arcminute field of view, this instrument was used for broad-band characterization and selection of AGN. The Multiband Imaging Photometer for Spitzer (MIPS) instrument[26] covered the mid- through far-infrared range. In particular, the 24 $\mu$m detector provided longer wavelength mid-infrared data on AGNs to complement the IRAC wavelengths (emission seen in the MIPS 70 $\mu$m and 160 $\mu$m bands was typically dominated by star formation, even in the hosts of AGNs). Perhaps the most critical instrument for studies of AGNs with Spitzer was the InfraRed Spectrograph (IRS).[27] This spectrograph covered 5-38 $\mu$m allowing for both low resolution (resolving power, R~60-100) and higher resolution (R~600) spectra (comparable to the SWS and LWS instruments on ISO). The instruments on Spitzer were able to make use of larger format and more sensitive array detectors than previous missions, making both imaging large fields, and obtaining spectra over a wide wavelength range much more efficient. The largest array on ISO (the ISOCAM imager) was 32x32 pixels, that on Spitzer (the IRAC camera) was 256x256 pixels. This, combined with a slightly larger telescope aperture (85 cm, giving a diffraction-limited resolution of 7 arcseconds at 24 $\mu$m, compared to a 60 cm aperture for IRAS and ISO and a 10 arcsecond diffraction limit at 24 $\mu$m), enabled highly efficient imaging and spectroscopy at angular resolutions of a few arcseconds.

In this review, we begin by discussing how Spitzer spectroscopy gave us insights into the geometric distribution and nature of the dust obscuring the central AGN including the role of the host galaxy therein. We then discuss how Spitzer data were utilized to isolate objects with the characteristic spectra of AGNs at these wavelengths to select new samples of AGNs out to redshifts z~4. These samples are much less affected by dust reddening and extinction than other optical techniques, and unaffected by gas columns that cause absorption in the X-ray. Studies of AGN and quasar evolution based on these samples are then described, and their implications for the nature of the obscured population discussed. Spitzer studies in the context of AGN selected at other wavelengths are then briefly reviewed. Finally, the contribution of Spitzer to the understanding of the host galaxies of AGNs and the coevolution of black holes and galaxies is discussed, and the contribution of Spitzer to studies of AGNs, past and future, assessed. Throughout this review the following conventions are adopted: near-infrared is used to describe the wavelength range from 0.9 - 5 $\mu$m; mid-infrared is used to describe the range 5 - 30 $\mu$m; and quasars refer to AGN with bolometric luminosities $>\sim 10^{12}$ L$_\odot$ ($10^{45.6}$ erg s$^{-1}$).

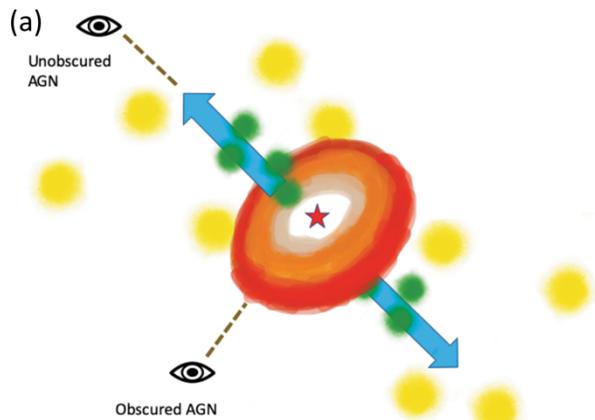
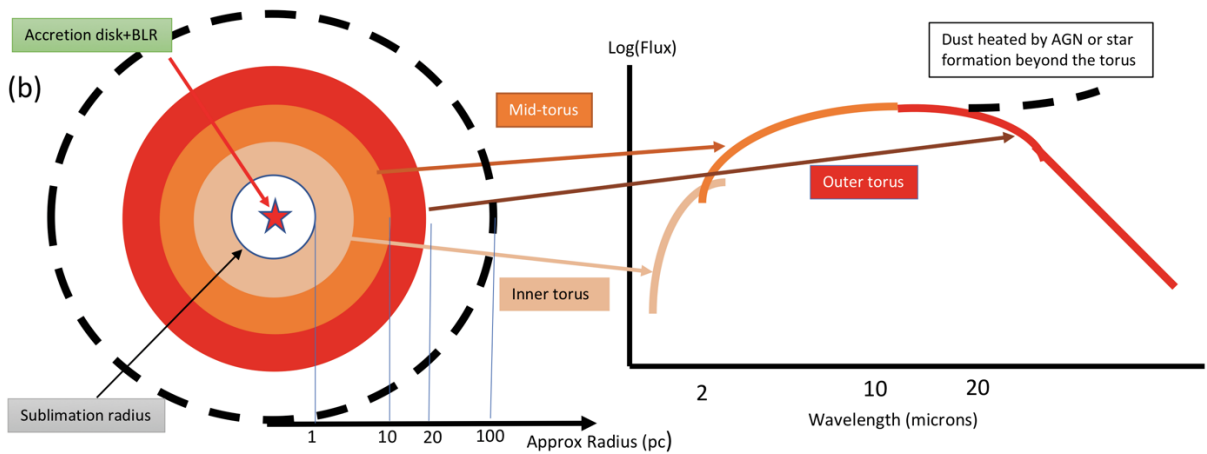

Figure 1: The torus model has been very successful at both explaining why AGN are seen directly in some objects and not in others, and also for predicting the mid-infrared emission from AGN. (a) An overview of the model, with the central AGN indicated by the red star and surrounding torus in shades of red, showing that an unobscured AGN is seen when the torus is observed face on, and an obscured one when observed edge-on. Outside of the torus region, emission from warm dust in narrow-line emitting clouds (yellow) and polar dust clouds (green), perhaps originating in a wind (blue), can also contribute to the dust emission, particularly at longer mid-infrared wavelengths. (b) The torus emits at infrared wavelengths, starting at about 1 μm with the hottest at the dust sublimation radius ~1 pc and extending out to ~20 pc for a luminous AGN (bolometric luminosity ~$10^{46}$ erg s$^{-1}$), with the mean dust temperature falling with radius (the flux axis in panel (b) is in arbitrary units). Emission from the torus dominates the SED out to 20 μm [ref.17]. Below, we indicate how different regions of the torus contribute to the overall AGN mid-infrared SED.

**Mid-infrared spectroscopy of AGNs**

Spitzer's IRS instrument allowed astronomers to perform the first comprehensive studies of the mid-infrared spectra of relatively local, previously known AGNs including: optically-selected Type-1 and Type-2 AGNs from Seyferts to quasars,[28-32] X-ray selected AGNs including Compton-thick AGNs,[33] low-luminosity AGNs and radio AGNs.[34,35] These studies helped confront and refine models of the geometric distribution of the obscuring structure around the central AGN [e.g. 36] as well as the nature of the dust in the vicinity of AGNs. Below we highlight the key results of this work.

As expected, based on ISO results, the IRS instrument on Spitzer detected narrow line high ionization emission lines in the mid-infrared spectra of AGNs (Figure 2; panel (a)). In particular, the [NeV] emission line at 14.3 μm, [ArV] at 13.1 μm and the [OIV] emission line at 25.9 μm are very weak or non-existent in starbursts, and their presence in mid-infrared spectra is thus a reliable indicator of an AGN. Spitzer increased the samples with such detections by roughly an order of magnitude from the few tens found by ISO.[40] Spectral line diagnostics gave insights supporting models in which Seyfert 1 and Seyfert 2 galaxies are the same systems viewed from different angles.[41] In particular, the bright [OIV] 25.9 μm line is well correlated with the hard-X ray luminosity in AGNs, and thus may be a good, orientation-independent measure of intrinsic AGN luminosity.[42,43] IRS spectra have also been used to confirm the presence of AGNs in ULIRGs via [NeV] emission in cases where other techniques have proved ambiguous.[44] It was found, on the basis of the Spitzer IRS spectra of known AGNs with reverberation mapping-based black hole mass estimates, that the velocity dispersions derived from these high ionization lines (specifically [NeV] and [OIV]) correlate with the black hole masses. This paves the way to measuring black hole mass based on mid-infrared spectra alone -- a technique that is sure to be very useful in the future with the JWST.[45]

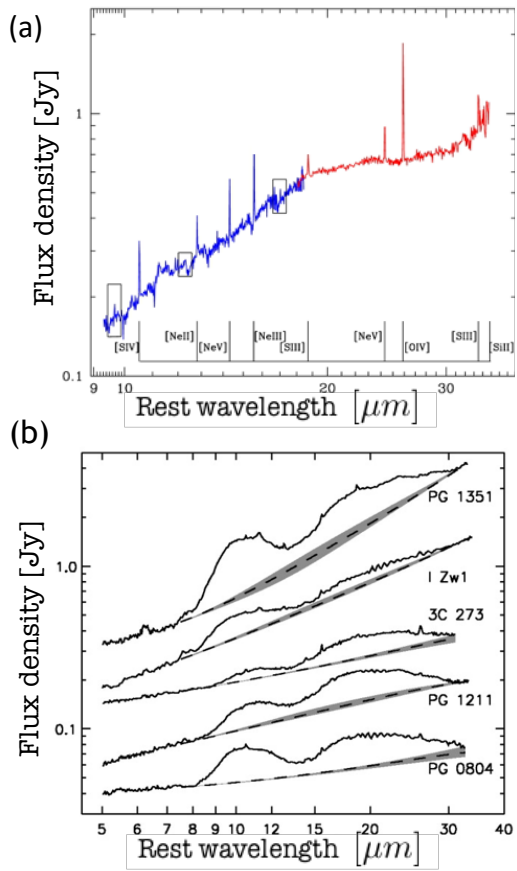

**Figure 2: Spectroscopic diagnostics of AGN in the mid-infrared.** (a) A typical unobscured/Type 1 AGN mid-IR spectrum showing no PAH features, but prominent fine structures lines due to high ionization species such as [NeV]14.3 µm and [OIV] (adapted from ref. [37]; the blue spectrum is from observations in the IRS short-high module and the red those from the long-high module ). (b) In support of the orientation/torus model, Spitzer IRS spectra of unobscured quasars for the first time showed the silicate features at 9.7 and 18 µm in emission (adapted from ref. [38]). (c) In support of the evolutionary scenario, deep silicate absorption features were found to correlate with strong host galaxy obscuration especially in the compact coalescence stage of a major merger. This led to a diagnostic plot based on the PAH equivalent width (proxy for the ratio of AGN to star-formation) vs. the silicate feature depth. (adapted from ref. [39]).

Figure 2 (a) shows a particular, illustrative, mid-infrared spectrum for a Type 1 AGN. However, there is considerable spread in both the overall shape as well as the presence and strength of silicate absorption/emission features at 9.7 and 18 µm. Detailed radiative transfer models of the mid-infrared SEDs of AGN tori broadly fall into two categories: smooth, or clumpy, depending on whether the dust is modeled as a smooth component or in distinct clouds/clumps (see [46] for a comparison of the two). Detailed model fitting of such Spitzer mid-infrared spectra of Type 1 AGN suggests the presence of a more complex obscuring structure than a simple dusty torus. One approach[47] uses a three-component model including a standard ISM clumpy torus, a pure-graphite hot dust originating in between the torus and BLR which dominates the near-IR emission, as well as emission from dusty clouds in the NLR which dominates the intrinsic AGN emission beyond 20 µm (see [36] for a review). Such models however are somewhat degenerate with the assumptions about the smoothness or clumpiness of the dust in the torus.[46] In addition, over the last decade, high angular resolution ground-based mid-infrared imaging has led to a model where the outer "warm torus" may actually be due to a dusty polar wind.[48-50] This scenario was tested against the large samples of AGNs that now have Spitzer constraints and found to be generally consistent with the diversity of mid-infrared SEDs seen therein.[51] Clearly, more complex multi-component models are needed to explain the intrinsic near- and mid-infrared emission of AGN, even before the role of the host galaxy is included.

Regardless of the specific dust geometry in the torus, a clear prediction of the orientation model (Figure 1) is that in obscured quasars/AGN the mid-infrared silicate features should be seen in absorption only, whereas for the unobscured quasars we could see the silicate features in emission. Spitzer IRS studies of previously-known unobscured quasars observed this emission feature for the first time[38,52-54] (Figure 2 (b)), a key piece of evidence showing a clear line of sight to hot dust close to the AGN, and consistent with the presence of a dusty torus or disk.

As expected, obscured AGNs (including Seyfert 2s and even more extreme Compton-thick AGNs, with neutral hydrogen column densities > $10^{24}$ cm$^{-2}$) nearly always[55] show the silicate feature in absorption.[41,54] However, this feature is typically not very deep in such sources, consistent with clumpy torus models (e.g. [15,46]) where only a modest optical depth can be built up.[33] The rare sources where very deep silicate absorption features are seen (Figure 2(c)) may either be closer to the smooth torus models or have additional obscuration due to host galaxy dust.

Much of the modelling literature treats the mid-infrared spectra of sources containing AGN as a sum of an AGN torus spectrum plus host galaxy/starburst spectrum. While a reasonable assumption for unobscured (Type 1) AGN where we know we have a direct line of sight to the accretion disc and surrounding torus, for obscured (Type 2) AGN this is not so clear and it is possible that the emission from the AGN (accretion disc+torus) is re-processed by the host galaxy dust as well.[56] Figure 3 (a) shows a simulation of this effect where the AGN infrared emission spectrum after vs. before processing by the host galaxy dust is redder, peaks at longer wavelengths and has a deep silicate absorption feature. This particular simulation is at the coalescence stage of a major merger.[57]

This more sophisticated approach to understanding AGN obscuration is in line with the scenario where obscured AGNs are seen in an early phase of their lifecycle, leading to a prediction that powerful quasars should be present in the coalescence stages of major mergers where the host galaxy obscuration is also significant.[6] Indeed, Spitzer IRS spectra revealed sources with very deep silicate absorption likely associated with ``buried'' AGNs.[58] The highest silicate absorption depths were observed to correlate with high inclination of the host galaxy (see Figure 3 (b)) or with disturbed merger-driven morphologies (including local ULIRGs).[33,58,59] Simulations also suggested that this feature is deepest in the coalescence stage of gas-rich major mergers.[56] It appears that these evolution-driven obscured quasars are particularly prominent at higher luminosities and are rarer at lower luminosities; for example, the IRS spectra of roughly half the AGN with higher mid-infrared luminosities show deep silicate absorption, but at lower mid-infrared luminosities these are rare.[60,61,62] The existence of these host-obscuration driven deep silicate absorption features also suggested a diagnostic for the presence and evolutionary stage of an AGN in the form of the PAH equivalent width vs. the silicate feature depth[39] (Figure 2 (c)).

The Spitzer IRS spectra have not only been used to test theories for the geometric distribution of the obscuring dust around the central AGN and role of host galaxy obscuration therein, but also to give insights into the possible effects of the AGN on the dust properties. Detailed modeling of the dust properties, in particular shifts in the wavelength of the 9.7 μm silicate feature, both in emission and absorption, indicated that AGNs may be affecting the chemical composition,[63] the dust grain size distribution,[64] and/or the porosity of the dust grains,[65,66] though radiative transfer effects may also influence the position of the silicate emission feature.[55]

The Spitzer IRS instrument crossed a new frontier by being the first instrument capable of obtaining mid-infrared spectra of distant galaxies. This included obtaining spectra for the only previously known population of distant dust-obscured galaxies (the sub-mm galaxies or SMGs[67]) as well as obtaining spectra of the large population of dusty IR-luminous galaxies Spitzer itself had discovered with its MIPS instrument. These studies allowed for mid-infrared spectroscopic diagnostics out to cosmic noon (the epoch of peak build-up of stars and supermassive black holes in the Universe at z~2) for hundreds of galaxies.[61,68-73] They served several key purposes -- obtaining spectroscopic redshifts for dusty galaxies whose optical spectra are very challenging to observe, giving clues to the composition of the ISM at these early epochs,[74] and allowing the determination of the role of AGN heating in these dusty systems. These spectra have been an invaluable resource to correctly identify the AGN content not only for Spitzer (e.g. 24 μm) selected galaxies,[58] but also for populations found post-Spitzer, such as Herschel 250 μm-selected galaxies.[75]

These Spitzer IRS studies of large samples of IR-selected dusty galaxies (regardless of whether they contain a previously known AGN or not) led to the realization that for a complete census of AGN activity (aka census of supermassive black hole growth), we should not only be counting the systems where the AGN dominates, but also systems where there is both significant AGN and star-formation activity – i.e. composite sources.[59,61] Spitzer IRS based SED templates that incorporate the continuum from pure AGN to pure star-formation as opposed to a

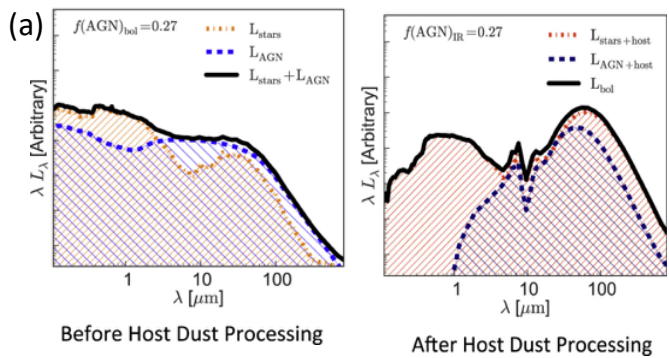

(a) 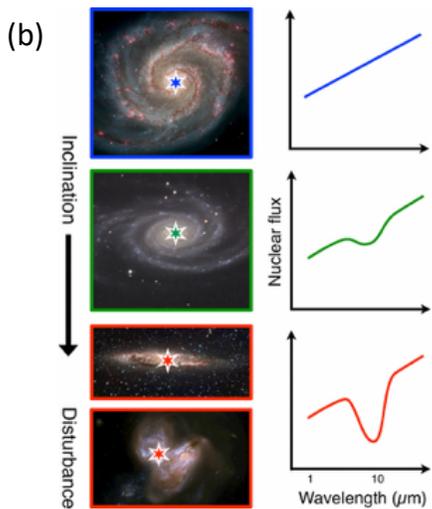 (b)

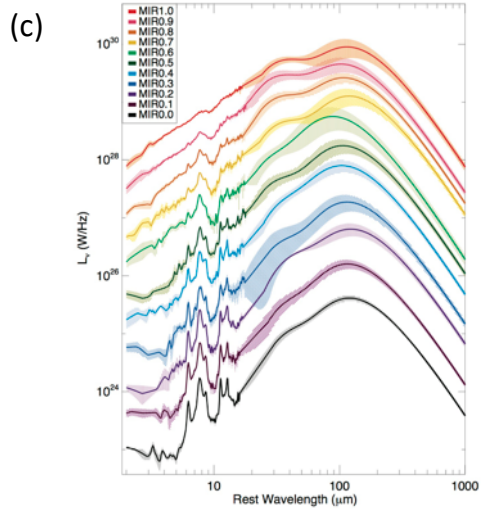 (c)

**Figure 3: The connection between an AGN and its host in the mid-infrared.** (a) In models when an unobscured AGN+torus model SED (blue dashed line in the left-hand panel) is embedded in a dusty host galaxy (red dot-dashed line), the mid-IR spectrum of the AGN is significantly altered (right panel). Dust processing changes the SED of the galaxy from the sum of the stars and AGN (black solid line in the left-hand panel). The AGN SED is reddened in the mid-IR, shows a deep silicate feature at 9.7 µm, and a strong far-IR component of emission, resulting in an SED that is the black solid line in the right-hand panel (adapted from ref. [57]). Not accounting for this longer wavelength emission from AGNs can strongly underestimate the contribution of the AGN to the overall power output of a galaxy. (b) This is supported by the Spitzer IRS spectra of Compton-thick AGNs, where the 9.7 µm silicate feature is not very deep. By contrast deep silicate absorption correlates with the inclination of the host galaxy.[33,58] It is also prominent in mergers (adapted from ref. [33]). (c) Near-IR through sub-mm SED templates showing the continuum of mid-IR spectra from pure star-forming galaxies (SFGs) to pure AGN. (adapted from ref. [59]).

binary split (Figure 3 (c)) help us account much more completely for the role of AGNs in the overall IR emission from galaxies. Indeed, objects where AGN and star formation activity coexist are the most important ones for understanding the role of AGNs in galaxy evolution.

**Selection of AGNs using Spitzer broadband colors**

The realization that mid-infrared colors could be used as AGN diagnostics began with ISO,[76,77] but it was only later with Spitzer's IRAC instrument that enough colors in the near- through mid-infrared range became available to greatly improve our ability to distinguish starbursts from AGNs. The largely featureless, red continuum emission from AGNs in the rest-frame 2 - 20 µm range[17] (see Figure 2 (a)) presents a unique broad-band spectral signature well suited to identification through simple color criteria [78-81] (Figure 4). The nature of the AGN luminosity function in the mid-infrared meant that shallow Spitzer surveys were most efficient at detecting AGN per unit telescope time invested. Integrating the luminosity function of [84] there are about 400 AGN/deg$^2$ to a flux density limit of 0.4 mJy at 24 µm. About 55 deg$^2$ were surveyed to approximately this depth in the cryogenic *Spitzer* mission in the SWIRE,[83] AGES,[85] First Look[86,87] and S-COSMOS[88] surveys, resulting in ~20,000 AGNs. The vast majority of these lack spectroscopic redshifts due to the difficulty of obtaining sufficiently deep spectra on existing optical/near-infrared telescopes.

The completeness and reliability of the AGN selection varies according to the exact technique used.[89-91] In general, simple color-based techniques tend to fail at high redshifts (z>1.5), where the redshifted colors of the stellar populations of normal galaxies can cause them to fall into the selection regions ("wedges") of [78,79] (hereafter the "Lacy" and Stern" wedges, respectively), [81] especially If the stellar SEDs are reddened by dust. By including the requirement of a detection at 24 $\mu$m, quiescent galaxies can be eliminated,[92,93] but high redshift star-forming galaxies are still a contaminant. Composite AGN/starburst objects will be missed at z<0.5 if the 7.7 $\mu$m PAH feature is strong enough to significantly redden colors in the IRAC 8 $\mu$m band, and low accretion rate objects (with Eddington ratios $\sim< 0.01$) that lack an accretion disk to heat dust close to the nucleus will also be missing.[81,93] Completeness and reliability of color-based AGN selection thus depends on the luminosity of the underlying AGN, with both completeness and reliability being higher for high luminosity AGN, when the AGN power-law dominates over emission from the stellar population and any dust heated by star formation.[90] Based on modeling and analysis of X-ray detected fraction, the Lacy wedge is less reliable than that of Stern wedge,[81] but more complete.[90] Optical/near-infrared spectroscopic follow-up of 786 AGN candidates selected using a slightly modified Lacy wedge, together with a requirement for a 24 $\mu$m detection, resulted in 527 AGN confirmations based on rest-frame optical/UV criteria, indicating a reliability in selecting AGNs of at least 67%, and perhaps higher if many optical/UV AGN diagnostic emission line were obscured, or diluted in composite systems.[93] The most reliable (but least complete) selection method of Donley et al.[81] is close to 100% reliable, so far as can be assessed from X-ray data. Thus, although these techniques are far from perfect, particularly for composite star-forming/AGN objects, and in deep surveys where stellar emission from high redshift galaxies is seen, their simplicity and ease of use have made them an important tool for AGN selection and classification.

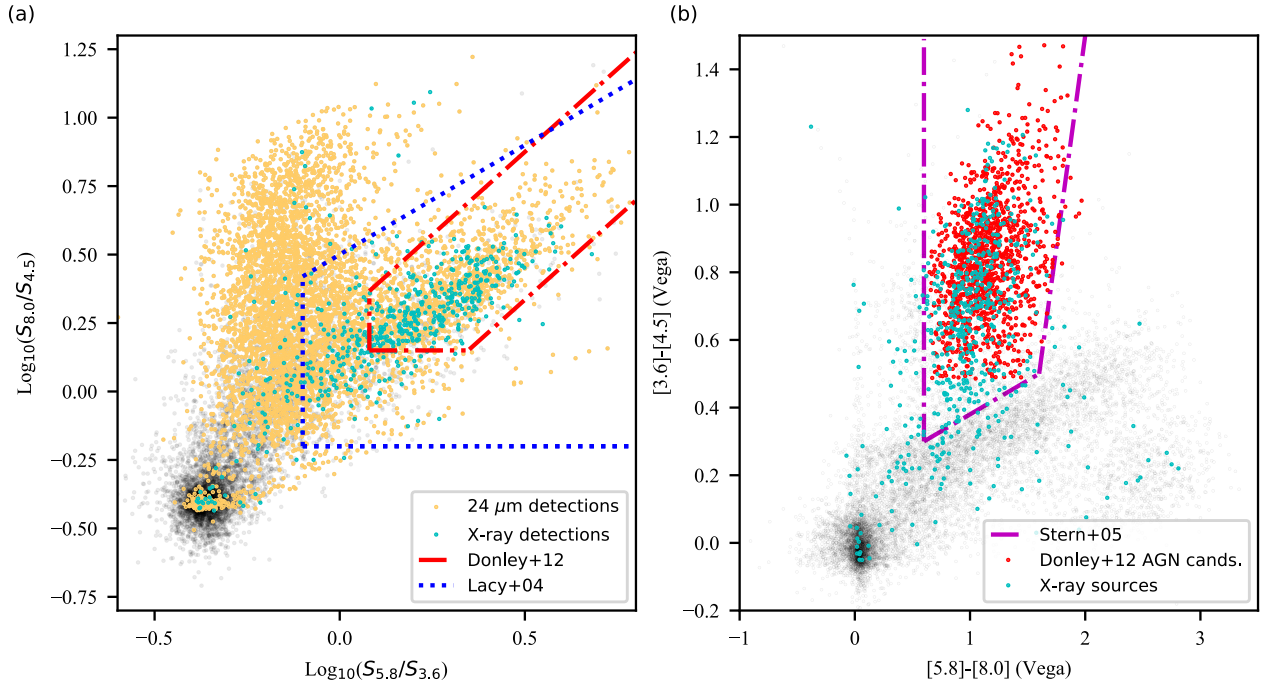

**Figure 4: Comparison of mid-infrared AGN selection techniques to results from the X-ray.** The cyan dots in both panels represent the 972 X-ray sources from the X-SERVS survey[82] with 4-band IRAC detections in SWIRE.[83] (a) The criteria of Lacy et al.[78] (blue dotted line) and Donley et al.[81] (red dot-dashed line) are shown. The background grey dots are the overall population of IRAC sources and the orange dots are IRAC sources detected at 24 $\mu$m. (b) The criteria of Stern et al.[79] (dot-dashed purple line) are shown, with 1175 AGN candidates from the highly-reliable selection of Donley et al. that are not detected in the X-ray by X-SERVS shown as the red dots (1947 and 3092 objects undetected in the X-ray satisfy the less reliable, but more complete criteria of Lacy et al. and Stern et al., respectively).

Optical/infrared spectroscopic follow-up of obscured AGNs selected using IRAC colour criteria showed that they find not only the expected population of Type-2 AGNs and quasars, but also a significant population of objects that are Type-1 quasars reddened by dust, with E(B-V) color excesses compared to normal quasars of

~0.3-1.[92,93] This class of objects were known from follow-up of radio sources,[94-96] but radio-quiet equivalents had not been found in significant numbers until the advent of AGN surveys with Spitzer. Even these relatively small amounts of reddening and corresponding extinction are sufficient to remove them from optically-selected quasar samples, and they also have significant X-ray absorption.[97] The obscuration in these objects could originate either in the nuclear region, or in the non-nuclear regions of the host galaxy.

Spitzer bands have also been used to improve colour-based selection of Type-1 quasars. By expanding the color space available to quasar selection from optical colours into the near-infrared, it is easier to distinguish quasars from stars whose optical colours can mimic those of high-z quasars.[98,99] This technique can work well using only the IRAC 3.6 $\mu$m and 4.5 $\mu$m bands, allowing selection of fainter quasars, and those observed in the post-cryogenic phase of Spitzer, when only 3.6 $\mu$m and 4.5 $\mu$m data were obtainable.[100]

**Luminosity/number density evolution**
Spitzer-selected samples of AGNs have been used to construct an AGN luminosity function.[84] Figure 5 shows the AGN luminosity function from a combination of Spitzer and Wide Field Infrared Survey Explorer data.[101] (As described in Section 6, WISE AGN surveys built upon the mid-infrared AGN selection techniques developed for Spitzer, but cover the whole sky to a relatively shallow depth, so are able to fill the high luminosity, low redshift population largely missing from the deeper, but smaller area Spitzer surveys.) This shows similar behavior to X-ray and optical studies overall, but with a couple of notable differences. At the faint end, it suggests more AGNs than seen in the X-ray, consistent with a population of weak, but very highly obscured AGNs missing from X-ray samples. At the bright end, although the Type-2 AGN number density is lower than that of the Type-1s at high luminosity, the number density of red and blue Type-1s is comparable in the highest redshift bin (z~2). At z>2, there is a suggestion that the obscured AGN population peaks in number density at higher redshift than the unobscured population,[84] similar to the radio-loud AGN population.[102] The fraction of Type-2 objects is seen to decrease with AGN bolometric luminosity, as seen in samples selected at other wavelengths, but the fraction of lightly-reddened Type-1 AGNs is roughly constant with luminosity, and it is this class of moderately obscured object that seems to be prevalent at high luminosities and high redshifts (though selection effects may also play a role as highly obscured Type-2 objects will be increasingly difficult to find at high redshifts, even in the infrared).

Mid-infrared Spitzer surveys reveal a higher radiative efficiency for BH accretion over the history of the Universe, and thus possibly a higher BH spin rate, than indicated by optical and X-ray luminosity functions. By comparing the luminosity density emitted by AGN through the history of the Universe to the mass density in supermassive black holes today, the mean radiative accretion efficiency ($\eta$) of AGN can be estimated. [103-105] The larger number density of AGN found in mid-infrared surveys implies a higher luminosity density of AGN emission than predicted from X-ray or optical luminosity functions and thus a higher mean value of $\eta$. Using the X-ray AGN luminosity density alone, values of $\eta$~0.15, significantly larger than the 0.05 value predicted from a Schwarzschild black hole, indicating spinning Kerr black holes (which have closer innermost stable orbits, and hence higher radiative efficiencies), are found.[105] The discovery of an even higher AGN luminosity density than inferred from the X-ray using mid-infrared selected AGN confirms that most black holes are spinning, and leaves little room for black hole growth through non-radiative or radiatively inefficient processes.[84]

**Multiwavelength comparisons**
Comparison of AGN samples selected in the mid-infrared to those selected at other wavelengths helps to inform the relative selection biases of different AGN search techniques.[81,90,91] Many candidate AGNs detected by Spitzer mid-infrared diagnostics are not identified as AGNs at other wavelengths in comparable on-sky integration times. Conversely, surveys with Spitzer/IRAC were extremely efficient at detecting AGNs (Figure 4). For example, 97% of the AGNs found in 5 ks exposures with the Chandra X-ray Observatory in the X-Bootes field were identified in just 90 s of integration using IRAC, with 65% falling within the AGN selection criteria of Stern et al.[106] In the optical, AGNs may be missing due to obscuration of the optical lines, or the presence of strong emission from star formation in the host disrupting the standard AGN emission line diagnostics.[90] When compared to X-ray studies, Spitzer was able to identify luminous ($\nu L_\nu(6\ \mu m) > 6 \times 10^{44}$ erg/s/cm$^2$) Compton Thick quasars at z~2 that were not detectable in X-rays in 2-4 Ms observations with the Chandra X-ray Observatory, resulting in an estimated 24-48% fraction of such quasars that are Compton Thick,[107] consistent with recent estimates based on surveys with Chandra (e.g. [108] estimates a Compton Thick fraction of $38^{+8}_{-7}$%). (We assume a Compton Thick fraction of 1/3 for the remainder of this review.) Luminous Type-2 quasars were also identified from their ultraviolet spectra in the Sloan Digital Sky Survey, when followed up with Spitzer, these were shown to have continuum SEDs in the mid-infrared very similar to obscured AGNs selected in the mid-infrared, with a large range in PAH and silicate properties, but dominated by a strong continuum from hot and warm dust.[109]

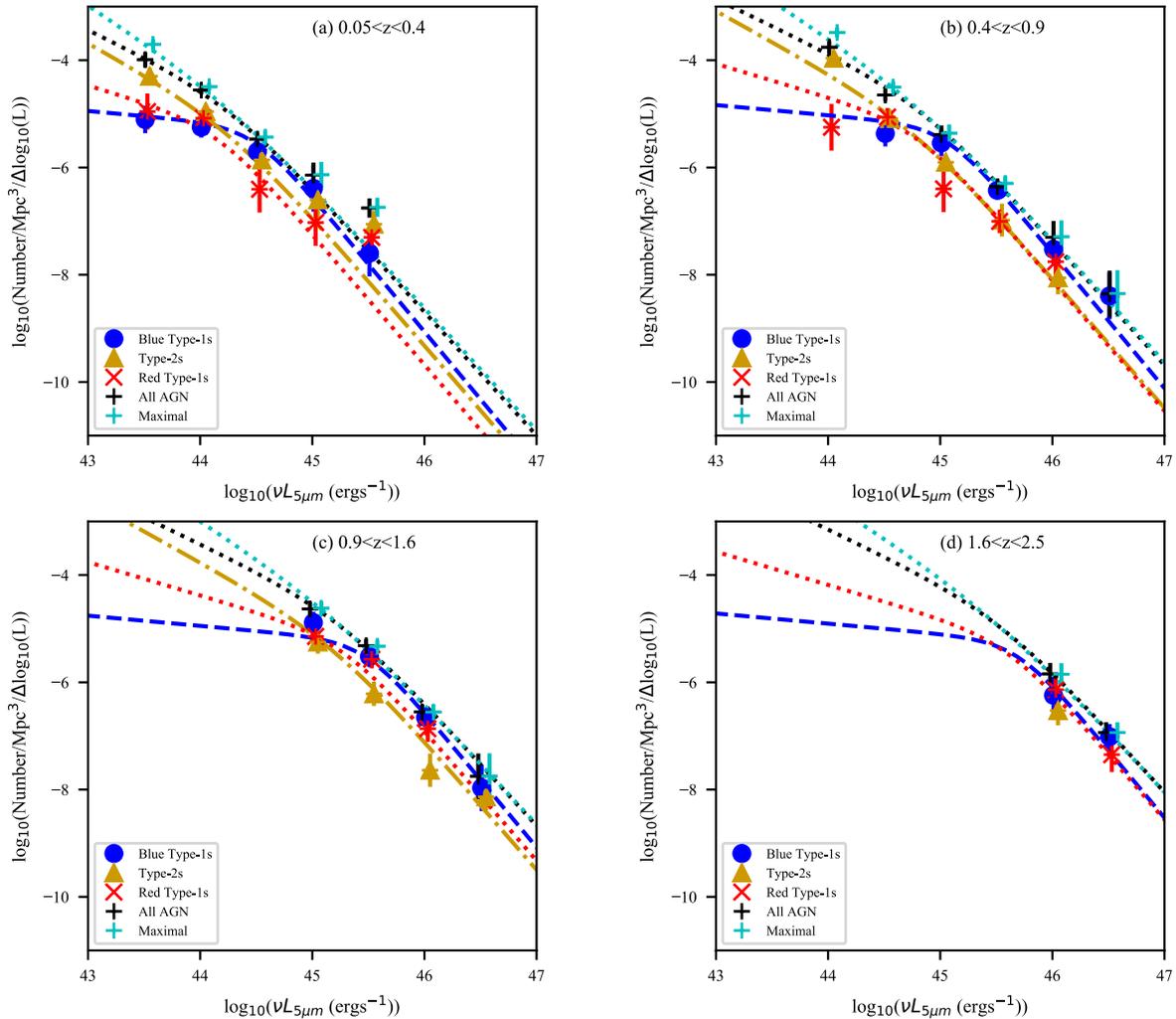

**Figure 5: The mid-infrared luminosity function of AGNs from the combined Spitzer and Wide Field Infrared Survey Explorer (WISE) sample in ref. [101].** The luminosity function at redshifts (a) 0.05-0.4; (b) 0.4-0.9; (c) 0.9-1.6, and (d) 1.6-2.5 plotted separately for blue Type-1 (blue circles/blue dashed line), red Type-1 (red diagonal crosses/red dotted line), Type-2 AGNs (orange triangles/orange dot-dash line) and the total AGN population (black crosses/black dotted line), along with a "maximal" estimate (cyan crosses/cyan dotted line) that includes objects selected as AGNs in the mid-infrared, but for which optical/IR spectroscopic confirmation was not possible. The points show the binned luminosity functions (with error bars), the lines the maximum likelihood fits.

Quantitative comparisons with surveys at other wavelengths are complicated by the difficulty of estimating accurate bolometric luminosities for AGN detected in the infrared. In the case of unobscured quasars, this is straightforward as it can be assumed that any infrared emission is from reprocessed optical/UV light, so can be ignored when making a bolometric correction. In the case of obscured quasars, we need to use the infrared to infer the intrinsic strength of the optical/UV continuum that is powering the infrared emission. We can use spectral energy distributions derived from multiwavelength observations of Type-1 AGNs and quasars to approximate this,[110] however, this correction, and its dependence on AGN luminosity, remains uncertain.

One class of AGN that is seen across a wide range of accretion rates, and which can be easily identified from surveys is radio-loud AGNs. Indeed, radio sources, either powered by AGNs or by starbursts, have very high detection rates (>95%) in Spitzer/IRAC data[111,112] (as indeed do X-ray sources[82]). Diagnostics from mid-infrared multi-band photometry can separate high accretion rate AGNs from starbursts.[113,114] The class of low accretion rate ($\sim$<1% of the Eddington Limit) radio sources, however, shows no detectable warm/hot dust emission in the mid-infrared and also typically lack high ionization emission lines both in the optical and infrared.[34] Instead, their detections are based on stellar photospheric emission from their massive host galaxies seen by IRAC. In general, Spitzer studies of known optically-selected AGNs with low accretion rates have little or no warm dust emission. Theoretical predictions[115-117] suggested that a certain minimum AGN luminosity (possibly also a function of black hole mass[118]) is needed to drive the disk wind from which the torus forms. Using a spectral decomposition of archival IRS spectra, [119] showed that the hot dust emission from the torus indeed disappears for AGN bolometric luminosities less than about $10^{42}$ erg s$^{-1}$.

**Co-evolution of AGNs and their host galaxies**

It has been noted for some time that the cosmic evolution of quasar and star-formation luminosity densities are very similar (at least for the unobscured populations of quasars and star-forming galaxies),[120,121] consistent with, but, of course, not proving, a causal link between the two processes. Such co-evolution may be driven by common fuel (gas supply) and mutual regulation through various feedback mechanisms. This co-evolution may be taking place in non-interacting galaxies driven by secular processes, or it may be associated with major mergers (leading to the evolutionary scenario for obscured quasars discussed in the Introduction).

Therefore, a key question in understanding this co-evolution is to understand the relative role of secular processes versus. mergers. Observations of host galaxies of the most highly obscured AGNs seem more consistent with these objects being seen at an early stage in their life cycles, with both objects selected in Spitzer[58,122] and X-ray surveys[123] being frequently seen to have evidence of mergers or interactions. In a study based on the HST CANDELS survey, [124] found that the luminous, heavily obscured AGNs that are only found in Spitzer mid-infrared surveys are significantly more likely to be in host galaxies that show signs of interactions or mergers. These observations are thus consistent with at least some component of the obscuration in the most highly-obscured objects arising in the host galaxy, rather than being confined to the few-pc scale of the torus, as discussed in Section 1 above (see also [10] for work prior to Spitzer). However, the host galaxies of the majority of AGNs detected in the optical or X-ray show no obvious signs of mergers or interactions,[125-127] indicating either a delay between any merger and AGN activity, or that the AGN can also be triggered by processes within the nucleus of host itself. The mass of dust in the torus itself can be very low ($\sim 10^5$ M$_\odot$ [ref. 128]), so the presence of a torus does not require that the host itself be extremely dusty. A study involving both X-ray selected and Spitzer selected AGN samples found evidence that the fraction of host galaxies that are mergers is a steeply increasing and redshift independent function of AGN luminosity, exceeding 50% for AGN with $L_{bol}$>10$^{45}$ erg s$^{-1}$, and that mergers account for at least 50% of the cosmic growth in black hole mass.[129] Secular processes dominate at lower AGN luminosities which accounts for the bulk of AGN by number. It is likely therefore that the peaks in both black hole accretion rate and stellar mass build-up at cosmic noon is likely a combination of increased gas supply and increased frequency of mergers.

Further strong evidence for co-evolution of galaxies and black holes is the M-sigma relation (see [130] for a review) which in essence represents a relationship between the cumulative past star-formation and black hole accretion histories of individual galaxies. Therefore, the next key question in understanding AGN-host galaxies co-evolution has to do with understanding how such a

> **The Legacy of Spitzer for studies of AGNs**
>
> - Spitzer surveys were very efficient at finding obscured AGN,[78,79] and led to a much more complete understanding of the demographics of the total AGN population, finding a Compton-thick AGN population of about 1/3 of the total AGN number density.[107] The high AGN luminosity density derived from the mid-infrared AGN luminosity function is consistent with most black hole growth being from radiatively efficient accretion onto spinning black holes.[84]
> - Spitzer spectroscopy made the first detections of silicate features from hot dust in emission, indicating a clear line of sight to hot dust in Type-1 AGNs, consistent with the dust being in a disk or torus around the AGN. [38,52-55]
> - Spitzer mid-infrared spectroscopy alone was shown to be sufficient to quantify the ratio of star formation to AGN activity in galaxies.[59]
> - Spitzer spectroscopy found evidence for shocked $H_2$ gas in the host galaxies of AGNs, indicating that feedback of the AGN activity onto the ISM of the host is significant.[133-136]
> - The discovery by Spitzer spectroscopy of deep silicate absorption in some AGNs that correlated with host galaxy disk orientation confirmed that AGN obscuration on galactic scales is common.[33,58]

correlation arises. For example, for merger-driven AGN, is the growth in stellar and black hole mass truly coincident or is one lagging behind the other? A joint Spitzer-Hubble Space Telescope study of dust reddened quasars suggested that the black hole growth seems to lag the growth of the host galaxy in these objects.[131] For the more common, secular process triggered AGN we expect multiple stochastic AGN episodes throughout the history of a galaxy as opposed to a single strong burst (as in the merger-driven scenario). A Spitzer-based study of sources selected at 24 $\mu$m in the COSMOS survey[132] showed that the balance between star formation and AGN activity at z~1 was consistent with the measured black hole mass - bulge mass relation provided the AGN duty cycle (fraction of time the AGN is active) was about a factor of 3-5 shorter than that of the starburst.

As mentioned above, co-evolution is believed to involve various feedback mechanisms that help regulate both star-formation and black hole growth. Feedback effects of the AGN on star formation were also studied with Spitzer, though its low resolution limited what was achievable. One notable result was that IRS spectroscopy of radio-loud AGNs resulted in the detection of extremely strong $H_2$ emission lines in the mid-infrared, indicative of shocks in the ISM probably induced by the radio jets.[133-135] These AGNs have high masses of molecular gas, but low rates of star formation, consistent with these shocks heating the ISM.[136]

**Conclusions and later work**

Spitzer surveys, especially wide-field, shallow surveys like SWIRE,[83] proved to be extremely efficient at selecting AGNs, both ones that could be identified via other means in the optical and/or X-ray, and also finding a significant population of AGNs at or near the Compton Thick limit in a few minutes of exposure time that were undetectable, or only barely detectable in Ms exposures in the X-ray. Over 20000 AGN candidates exist in the Spitzer archive, several thousand of which have been confirmed spectroscopically.[85,93,100] Estimates of the AGN luminosity function in the mid-infrared implied a high fraction of obscured accretion onto black holes (>~50%) and a high total luminosity density from AGNs.[84] The WISE mission, a 0.4 m space-based telescope that surveyed the entire sky at 3.5, 4.6, 12 and 22 $\mu$m, has built directly on this Spitzer Legacy of infrared AGN selection,[137] and produced all-sky catalogs of millions of AGN candidates selected in the mid-infrared.[138] The all-sky capability of WISE enabled it to find very rare objects, including populations of highly-obscured, highly-luminous quasars.[139,140] The AKARI mission, launched by the Japanese Space Agency in 2006, with a 0.67 m mirror, provided photometry in nine near- through mid-infrared bands, enabling refinement of color-based AGN selection.[141] It also had a 2.5-5 $\mu$m spectroscopic capability that Spitzer lacked, allowing detection of the 3.3 $\mu$m PAH feature in AGN as a further diagnostic of star formation in AGN host galaxies,[142] and rest-frame optical spectroscopy of bright high redshift quasars.[143,144]

Spitzer was key to confirming many aspects of the accretion disk plus torus picture for AGNs, and for showing that both go away at low accretion rates. Recently, the Atacama Large mm/submm Array (ALMA) has been able to deliver images of local AGNs at resolutions of a few pc in the submm band, indicating molecular circumnuclear disks in nearby AGNs, though in many cases they differ from the simple predictions of the torus model, with some disks being asymmetric, lacking holes, or including counter-rotating components,[145-147] in line with some theoretical predictions.[148]

Spitzer made important contributions to our understanding of the nature of obscured AGN. The ability to easily select samples of obscured AGNs and quasars[78,79,81] allowed follow-up observations of host galaxies that showed that infrared-selected obscured AGNs tend to be found in systems that are actively interacting or merging, where obscuration on all scales, from the pc-scales of the nuclear region to the kpc-scales of the host galaxy can exist.[33,58] These observations added considerable weight to the argument that AGN obscuration, although it can happen at any stage in the life cycle of an AGN, is more commonly associated with early stages, particularly in high luminosity objects at z>~0.5. Spitzer observations helped define many other issues, including the question of AGN and host co-evolution and the effects of shocks induced by AGN jets and/or winds on the interstellar medium of the host galaxy. The Herschel infrared space telescope, launched in 2009, opened up the far-infrared window, from about 100 $\mu$m, and was used to study the star formation in large numbers of quasars at z~2, with results that indicated that unobscured quasars had similar star formation rates to normal galaxies in the same mass range,[149] although a joint Herschel/ALMA study over the redshift range 0.5-4 indicated that quasars may have a different distribution of star formation rates than normal galaxies, with a significant fraction below the "main sequence" of star formation versus stellar mass.[150] However, a Herschel-based study also suggested that the star formation rate from obscured quasars selected in the mid-infrared was higher than in their unobscured counterparts.[151] Finally, there remains the question of the completeness of the AGN census. Even Spitzer AGN surveys needed AGN light to be able to propagate from the nuclear regions at 24 $\mu$m or shorter wavelengths in order for them to be detected. It is possible a sizable population of very deeply buried AGN could still exist that would need longer wavelength surveys to identify, possibly combined with far-infrared emission line diagnostics to separate them from the star-forming population that dominates the counts at >~ 60 $\mu$m.

Three future space missions currently planned or under construction will define the future of AGN studies in the infrared in the coming years. The James Webb Space Telescope (JWST) will cover the wavelength range 0.6 $\mu$m to 28 $\mu$m and, in the mid-infrared, will provide much higher angular resolution (~0.1"-0.5") and much better sensitivity than Spitzer over an instrument-dependent field of view of 2-10 arcmin$^2$. It will also restore the capability to take mid-infrared spectra of AGNs, with resolving powers ($R$) up to 3500. JWST will be able to undertake high resolution mid-infrared studies of nearby AGN at a resolution ~100 pc, locating and characterizing warm dust in the nuclear region and shocked gas in the vicinity of the AGN. JWST will also undertake very deep (though small area) surveys in the mid-infrared, identifying obscured AGN to very low luminosities at high redshifts, accurately measuring the faint end of the AGN luminosity function in the mid-infrared.[152] In the near-infrared, the upcoming SPHEREx mission[153] will survey the whole sky between 0.75 and 5 $\mu$m, providing low resolution ($R \sim 40 - 150$) spectra for ~ 1 billion galaxies and AGN. Finally, the SPICA mission[154], if approved, will cover the wavelength range 12 $\mu$m to 230 $\mu$m at $R \sim 300 - 25000$ and slits up to 10 arcmin in length, allowing emission line surveys of AGN and their host galaxies, in particular AGN diagnostic lines such as [OIV]25.9 $\mu$m and the atomic fine structure lines such as [OI] 63 $\mu$m that can give information on the photodissociation regions in the ISM. Looking beyond the next decade, the Origins Space Telescope mission concept[155] plans to deliver even more sensitivity and resolution than SPICA in the far-infrared.

Future ground-based observations in the near through mid-infrared with future 30 m class telescopes and adaptive optics will offer even higher resolution imaging that JWST, ~0.1 arcsecond at 10 $\mu$m (though will be 1-2 orders of magnitude less sensitive).[156] At longer mid-infrared wavelengths, the 6.5 m University of Tokoyo Atacama Observatory will give ground-based access to the 25-40 $\mu$m window for the first time.[157] In the mm regime, the next generation Very Large Array (VLA), a rebuild of the current VLA with more collecting area (allowing continuum sensitivities <1 $\mu$Jy/beam) , longer baselines (allowing resolutions of milliarcseconds) and frequency coverage up to 115 GHz, will also be a powerful instrument for high resolution studies of AGN gas, dust and synchrotron emission on pc scales, and of radio-jet induced feedback in AGN at higher redshifts.[158] ALMA too will likely be expanded to increase its sensitivity by a factor ~10, with a wider bandwidth and more collecting area,[159] allowing deeper, higher resolution studies of dust continuum and molecular gas emission in AGN. The combination of new facilities is sure to both answer many of our existing questions about the nature of AGNs and their role in galaxy formation, as well as cause us to ask new ones.

### Acknowledgments

We would like to thank the dedicated staff at the Spitzer Science Center and the members of the Instrument Teams who were invaluable in making Spitzer the highly successful mission it was. We thank the referees for their thorough readings of the draft and many helpful comments.The National Radio Astronomy Observatory is a facility of the National Science Foundation operated under cooperative agreement by Associated Universities, Inc.

### Author Contributions
Both authors contributed equally to the planning and writing of this article.

### Competing Interests
The authors declare no competing financial interest.

Correspondence should be addressed to M.L. (mlacy@nrao.edu)